\documentclass[a4paper, 11pt]{article}
\usepackage[english]{babel}
\usepackage{amsmath, amssymb, amsfonts, bm}
\usepackage{mathrsfs}
\usepackage{textcomp}
\usepackage[mathcal]{euscript}
\usepackage{enumerate}
\usepackage{mathbbol}
\usepackage{float}
\usepackage{color,graphicx,epstopdf}
\usepackage{dsfont}
\usepackage{hyperref}
\usepackage{caption}
\usepackage{subcaption}
\usepackage[noadjust]{cite} 
\usepackage{mathtools}
\usepackage{tikz-cd}
\usepackage{mathtools}
\usepackage{algorithm}
\usepackage{algpseudocode}
\usepackage{multirow}
\usepackage{authblk}
\usepackage{blindtext}
\usepackage{hyperref}
\usepackage{menukeys}
\renewmenumacro{\directory}{pathswithfolder}

\newcommand{\bfu}{{\mathbf u}}
\newcommand{\bfU}{{\mathbf U}}
\newcommand{\bfx}{{\mathbf x}}
\newcommand{\bfX}{{\mathbf X}}

\newcommand{\bff}{{\mathbf f}}

\newcommand{\mcalV}{{\mathcal V}}
\newcommand{\mcalT}{{\mathcal T}}
\newcommand{\mcalS}{\mathcal S}
\newcommand{\mcalU}{{\mathcal U}}

\newcommand{\RICEGAME}{{\em RICE-GAME} }

\usepackage[left=1.5cm,right=1.5cm,top=2cm,bottom=2cm]{geometry}
\usepackage{float}

\title{\bf A Matlab and CasADi-based Implementation of \\ RICE Dynamic Game}
\date{}

\author{Yijun Chen}
\author{Guodong Shi}
\affil{\em The  Australian Center for Field Robotics, School of Aerospace, Mechanical and Mechatronic Engineering, The University of Sydney, NSW 2006, Australia, email: \{yijun.chen, guodong.shi\}@sydney.edu.au} 

\begin{document}
	\maketitle

\begin{abstract}
The most widely used integrated assessment model for studying the economics of climate change is the dynamic/regional integrated model of climate and economy (DICE/RICE) \cite{DICE,RICE}. In this document, we first represent the RICE-2011 model as a dynamic game, termed the RICE game. Then, both cooperative and non-cooperative solutions to the RICE game are considered. Next, a description of how to use the repository {\RICEGAME} on GitHub is provided \cite{RICE-GAME}. The repository \RICEGAME is a Matlab and CasADi-based implementation of the RICE game and its cooperative and non-cooperative solutions.
\end{abstract}

\section{Preliminary: Dynamic Games}\label{sec:DTDG}
	The theory of dynamic games lies in the interface between   game theory and optimal control, which involves a dynamic decision process for multiple players, each of which tries to maximize a cumulative payoff function. An $n$-player discrete-time dynamic game over a finite horizon is defined as follows. 

\medskip

{\noindent \bf Dynamic Game.} The $n$ players are indexed in $\mcalV:= \{1,2,\dots,n\}$; time is discrete with the steps  indexed in $\mcalT:= \{0,1,\dots,T\}$. Each player can manipulate the game through its control decisions, and the control decision space of player $i \in \mcalV$ is denoted by $\mcalU_{i} \subseteq \mathbb{R}^{d}$. At each time step $t=0,\dots,T$, the   decision executed by player $i$  is denoted  by $\bfu_{i}(t) \in \mcalU_{i}$. We also use $\bfu(t) = [\bfu_{1}^{\top}(t);\dots;\bfu_{n}^{\top}(t)]$, $\bfU_{i} = [\bfu_{i}^{\top}(0);\dots;\bfu_{i}^{\top}(T)]$ and $\bfU = [\bfU_{1};\dots;\bfU_{n}]$ to represent the all-player decision profile at time $t$, the player-$i$ decision throughout the time horizon, and the decision profile for all players and for all time stpes. The control decisions of all players excluding player $i$ at time step $t$ is denoted by $\bfu_{-i}(t)$, and the control decisions of all players excluding player $i$ over the entire horizon is represented  by $\bfU_{-i}$.

For each $t=0,1,\dots,T$, the group of players are associated with a dynamical state    $\bfx(t)\subseteq \mathbb{R}^{m}$ denoting the state space, whose dynamics are described by 
\begin{equation}\label{eq:DTDG_dynamics}
	\bfx(t+1) = \bff(t, \bfx(t),\bfu(t)), \quad \bfx(0) = \bfx_{0}, \quad t=0,\dots,T,
\end{equation}
with $ \bfx_{0}$ being the initial state. At each time $t=0,\dots, T$, upon playing $\bfu_{i}(t)$, the agent $i$   receives a payoff $g_{i}(\bfx(t),\bfu_{i}(t),\bfu_{-i}(t)) \in \mathbb{R}$ given other players' actions $\bfu_{-i}(t)$ and the current state $\bfx(t)$, where  $g_{i}(\bfx(t),\bfu_{i}(t),\bfu_{-i}(t))$ is a continuous function with respect to $\bfx(t)$,$\bfu_{i}(t)$, and $\bfu_{-i}(t)$. The cumulative payoff of agent $i$ throughout the time horizon is therefore 
\begin{equation}\label{eq:DTDG_payoff_function}
	\mathsf{J}_i(\bfX,\bfU_{i},\bfU_{-i}) = \sum_{t=0}^{T}g_{i}(\bfx(t),\bfu_{i}(t),\bfu_{-i}(t))
\end{equation}
where $\bfX=(\bfx^\top(0), \dots, \bfx^\top(T))^\top$. 
Each player's goal is to maximize its own payoff function by making best control decisions for the underlying dynamical system, and the system dynamics produces a terminal state $\bfx(T+1)$ towards the end of the time horizon. 

\section{Description of RICE as a Dynamic Game}
In what follows, we represent  the RICE-2011 model as a dynamic game, termed the RICE game. Our presentation of the RICE game is based on the RICE-2011 model with slight modifications, but the nature of being a dynamic game is preserved. 
\medskip

{\noindent  \textbf{Regions.}} There are $12$ regions in the RICE game. Each region is considered a player and the regions are  indexed in $\mcalV = \{1,2,\dots,n\}$ with $n=12$. 
\medskip

{\noindent \textbf{Time Steps and Calendar Years.}} The RICE game operates in periods of $\Delta = 5$ years, starting from the year 2020 as the initial year\footnote{The  RICE-2011 model operates in periods of $10$ years starting from 2005.}. Taking  the discrete time step index $\mcalT = \{0,1,\dots,T\}$, the relation between an actual calendar year and the corresponding discrete time step is determined by  
\begin{equation}\label{eq:time_step}
	year(t) = year(0) + 5t, \quad year(0) =2020,
\end{equation}
yielding calendar year $2020, 2025, 2030,\dots$ as desired.
\medskip

{\noindent \textbf{State Variables and Control Decisions.}} The RICE game has $n+5$ state variables: two variables to model the temperature dynamics in the form of the temperature deviation in the atmosphere and in the lower ocean from the reference year $1750$ ($T^{\rm AT}$ and $T^{\rm LO}$, respectively), three variables to model the carbon dynamics in the form of average carbon mass in the atmosphere, the upper ocean and biosphere, and the deep ocean ($M^{\rm AT}, M^{\rm UP}$ and $M^{\rm LO}$, respectively), and $n$ variables to model the economic dynamics in the form of each region's capital  ($K_{i}, i \in \mcalV$). Control decisions of the RICE game are each region's emission-reduction rate $\mu_{i}$ and each region's saving rate $s_{i}$ where the former represents the ratio of investment to the net economic output in each region, and the latter represents the rate at which ${\rm CO_{2}}$ emissions are reduced in each region.

We denote the dynamical  state of the RICE game at time step $t \in \mcalT$ as
\begin{align} \label{eq:rice_state_variables}
	\bfx(t) = [T^{\rm AT}(t); T^{\rm LO}(t); M^{\rm AT}(t);  M^{\rm UP}(t); M^{\rm LO}(t); K_{1}(t); \dots; K_{n}(t)] \in \mathbb{R}^{n+5}.& 
\end{align}
We also denote control decisions of region $i \in \mcalV$ at time step $t \in \mcalT$ by
\begin{equation}\label{eq:rice_control_i}
	\bfu_{i}(t) = [\mu_{i}(t);s_{i}(t)]^\top:=[\bfu_{i[1]}(t);\bfu_{i[2]}(t)]^\top \in [0,1]^{2}.
\end{equation} 
Consequently, control decisions of the RICE dynamic game at time step $t \in \mcalT$ of all players are 
\begin{equation}\label{eq:rice_control_whole}
	\bfu(t) = [\mu_{1}(t);\dots,\mu_{N}(t);s_{1}(t);\dots;s_{N}(t)] \in [0,1]^{24}.
\end{equation} 
Further denote region $i$'s control decisions over the time horizon by $\bfU_{i} := [\mu_{i}(0);\dots, \mu_{i}(T);s_{i}(0);\dots;s_{i}(T)]$.
The control decisions of all regions excluding region $i$ at time step $t$ is denoted by $\bfu_{-i}(t)$, and the control decisions of all regions excluding region $i$ over the time horizon is represented  by $\bfU_{-i}$.
\medskip
 
{\noindent \textbf{RICE Dynamics.}} The RICE game is also driven by  several exogenous and time-varying signals that evolve independently such as radiative forcing caused by greenhouse gases other than $\rm CO_2$ emissions $F^{\rm EX}$, each region's natural ${\rm CO_{2}}$ emissions due to land use $E_{i}^{\rm land}$, each region's carbon intensity $\sigma_{i}$, each region's total factor productivity $A_{i}$ and each region's population $L_{i}$. The dynamics of $\bfx(t)$ can be written as 
\begin{equation}\label{eq:dynamics}
	\bfx(t+1) = \bff(t, \bfx(t),\bfu(t)), \quad \bfx(0) = \bfx_{0}, \quad t \in \mcalT,
\end{equation}
where $\bfx_{0}$ represents the initial state of the underlying dynamical system. The transition functions $\bff:=[f_{1};f_{2};\dots;f_{n+5}]$ follow the RICE dynamics:
	\begin{align}\label{eq:temperature_dynamics}
	\begin{bmatrix}
		T^{\rm AT}(t+1) \\
		T^{\rm LO}(t+1)
	\end{bmatrix}
	&=
	\begin{bmatrix}
		\phi_{11}&\phi_{12}\\
		\phi_{21}&\phi_{22}
	\end{bmatrix}
	\begin{bmatrix}
		T^{\rm AT}(t) \\
		T^{\rm LO}(t)
	\end{bmatrix} 
	+
	\begin{bmatrix}
		\xi_{2}\\0
	\end{bmatrix}F(t),
	\end{align}
	\begin{align}\label{eq:carbon_dynamics}
	\begin{bmatrix}
		M^{\rm AT}(t+1) \\
		M^{\rm UP}(t+1) \\
		M^{\rm LO}(t+1)
	\end{bmatrix}
	&=
	\begin{bmatrix} 
		\zeta_{11}&\zeta_{12}&0\\
		\zeta_{21}&\zeta_{22}&\zeta_{23}\\
		0&\zeta_{32}&\zeta_{33}\\
	\end{bmatrix}
	\begin{bmatrix}
		M^{\rm AT}(t) \\
		M^{\rm UP}(t) \\
		M^{\rm LO}(t)
	\end{bmatrix} 
	+
	\begin{bmatrix}
		\xi_{1}\\0\\0
	\end{bmatrix} E(t),
	\end{align}
	\begin{align}\label{eq:captial_dynamics}
	K_{i}(t+1)&=(1-\delta^{\rm K}_{i})^{5} K_{i}(t) 
	+5\Big(1 - a_{i}^{[1]}T^{\rm AT}(t) - a_{i}^{[2]}T^{\rm AT}(t)^{a_{i}^{[3]}}\Big)\nonumber\\ &\cdot \Big(1-\theta_{i}^{[1]}(t) \mu_{i}(t)^{\theta_{i}^{[2]}}\Big)   A_{i}(t)K_{i}(t)^{\gamma_{i}}                                    \Big(\frac{L_{i}(t)}{1000}\Big)^{1-\gamma_{i}}s_{i}(t) ,
	\end{align}
where total radiative forcing  $F(t)$ at time step $t$, each region's total emissions $E_{i}(t)$ including natural emissions and industrial emissions at time step $t$ and global emission $E(t)$  at time step $t$ are given by
\begin{equation}\label{eq:rf}
	F(t) = \eta \log_{2}\Big(\frac{M^{ \rm AT}(t)}{M^{\rm AT,1750}}\Big) + F^{\rm EX}(t),
\end{equation}
\begin{equation}\label{eq:e_i}
	E_{i}(t) =  \sigma_{i}(t)(1-\mu_{i}(t))A_{i}(t)K_{i}(t)^{\gamma_{i}}(t)\Big(\frac{L_{i}(t)}{1000}\Big)^{1-\gamma_{i}} +E^{\rm land}_{i}(t) ,
\end{equation}
\begin{equation}\label{eq:e}
	E(t) = \sum_{i=1}^{n} E_{i}(t),
\end{equation}
\begin{equation}
	\theta_{i}^{[1]}(t) = \frac{pb_{i}}{1000\cdot\theta_{i}^{[2]}}(1-\delta^{pb}_{i})^{t}\cdot\sigma_{i}(t).
\end{equation}
 	\medskip
 	
 {\noindent \textbf{Payoff Functions.}} For each region, the social welfare at time step $t$ of the population $L_i(t)$ consuming the consumption $C_i(t)$ is defined by the population-weighted utility of per capita consumption 
 \begin{equation}\label{eq:utility}
 	g_{i}(C_{i}(t),L_{i}(t)) =  L_{i}(t) \cdot \frac{(\frac{C_{i}(t)}{L_{i}(t)}) ^{1-\alpha_{i}}-1}{1-\alpha_{i}},
 \end{equation}
where each region's consumption at time step $t$ is given by 
\begin{equation}\label{eq:compt}
	C_{i}(t) = \Big(1 - a_{i}^{[1]}T^{\rm AT}(t) - a_{i}^{[2]}T^{\rm AT}(t)^{a_{i}^{[3]}}\Big) \Big(1-\theta_{i}^{[1]}(t) \mu_{i}(t)^{\theta_{i}^{[2]}}\Big)   A_{i}(t)K_{i}(t)^{\gamma_{i}}\Big(\frac{L_{i}(t)}{1000}\Big)^{1-\gamma_{i}}s_{i}(t).
\end{equation}
  Each region's payoff function is defined as the cumulative social welfare of  region $i$ across the time horizon:
 \begin{align}\label{eq:social_welfare}
 	\mathsf{J}_i &=  \sum_{t=0}^{T} \frac{g_{i}(C_{i}(t),L_{i}(t))}{(1+\rho_{i})^{5t}} \nonumber\\
		&= \sum_{t = 0}^{T} \Bigg( \Big[ \frac{A_{i}(t)L_{i}(t)^{1+\alpha_{i}-\gamma_{i}}}{(1-\alpha_{i})(1+\rho_{i})^{5t}}\Big(1-\bfu_{i[2]}(t)\Big)\Big(1 - a_{i}^{[1]}\bfx_{1}(t) - a_{i}^{[2]}\bfx_{1}(t)^{a_{i}^{[3]}}\Big)\Big(1-\theta_{i}^{[1]}(t) \bfu_{i[1]}(t)^{\theta_{i}^{[2]}}\Big)\bfx_{5+i}(t)^{\gamma_{i}}  \Big]  
			\nonumber\\
			& \ \ \ - \frac{L_{i}(t)}{(1-\alpha_{i})(1+\rho_{i})^{5t}} \Bigg).
 \end{align}
For each region  $i \in \mcalV$, naturally it will attempt to maximize its cumulative social welfare. 
\medskip

We have now formally presented the RICE game where  the regions as players seek to  plan their control decisions in emission-reduction rate and saving rate for the entire time horizon  $$\bfU_{i} := [\mu_{i}(0);\dots, \mu_{i}(T);s_{i}(0);\dots;s_{i}(T)]$$ so as to maximize their payoff functions~\eqref{eq:social_welfare}  subject to the underlying dynamical system \eqref{eq:dynamics}, represented in \eqref{eq:temperature_dynamics}-\eqref{eq:captial_dynamics}.
\medskip

{\noindent \textbf{The Social Cost of $\rm CO_2$}.}  The regional SCC is then given by the ratio of the regional marginal welfare with respect to regional emissions and with respect to regional consumption
\begin{align}
	{\rm SCC}_{i}(t)&= -1000 \cdot \frac{\partial \mathsf{J}_i}{\partial E_{i}(t)}\Big/ \frac{\partial \mathsf{J}_i}{\partial C_{i}(t)} \notag \\ 
	&= -1000 \cdot \frac{\partial C_{i}(t)}{\partial E_{i}(t)} 	\label{eq:scc}.
\end{align}
\medskip

{\noindent \textbf{Parameters, Exogenous Signals and Initial State}.} The values for the parameters and initial state can be found in the tables at the end of this document. In the RICE game, the initial state are calibrated to match the data in the starting year $2020.$ The values for the parameters in the geophysical sector uses the latest updated values in the DICE-2016 model. The values for the exogenous and time-varying signals can be found in the Excel file named \path{exogenous_states_long.mat} in the repository \RICEGAME.
\section{Cooperative Decisions of the RICE Game}
In this section, we study the solutions to the RICE game under cooperative settings. First of all, we revisit the classical RICE solution concept defined by a system-level social welfare maximization. Next, we move to Pareto solutions to the RICE game. Finally, we introduce a receding horizon approach to approximate the classical RICE solution.
\subsection{Solution to RICE Social Welfare Maximization}
RICE social welfare maximization is for a centralized climate policy planner to compute the $\bfU_{i}, i \in \mcalV$ that maximize the sum of the weighed regional social welfare across all regions for a given initial condition $\mathbf{x}_0$ :
\begin{equation}\label{eq:swm}
	\begin{aligned}
		&  \max_{\bfU_{i}, i \in \mcalV}
		& &  \sum_{i = 1}^{n} c_{i} \mathsf{J}_i, \\
				&  {\rm subject \ to}
& & \bfx(t+1) = \bff(t, \bfx(t),\bfu(t)), \ \  \bfx(0) = \bfx_{0},  t \in \mcalT \\
&&& \bfu(t)   \in [0,1]^{24}, \ \  t \in \mcalT. 
	\end{aligned}
\end{equation}
where the values of $c_{i}, i \in \mcalV$ can be found in Table~\ref{table_ec_parameters}.

\subsection{Pareto Frontier between Developed and Developing Regions}
The regions in the RICE game are classified into two clusters of regions: developed regions (US, EU, Japan and other high income countries) and developing regions (Russia, Non-Russian Eurasia, China, India, Middle East, Africa, Latin America and other Asian countries). We denote $\mcalV_{\rm developed} = \{1,2,3,11\}$ and  $\mcalV_{\rm developing} = \{4, 5, 6, 7, 8, 9, 10, 12\}$.  Corresponding, the social welfare of the two clusters of regions are defined as, respectively, 
$$
\mathsf{W}_{\rm developed} = \sum_{i \in \mcalV_{\rm developed}} \mathsf{J}_i,\ \  \mathsf{W}_{\rm developing} = \sum_{i \in \mcalV_{\rm developing}} \mathsf{J}_i.
$$ 

We can calculate Pareto social welfare frontier between the developed and developing clusters  by solving the family of  optimization problems for a given initial condition $\mathbf{x}_0$:
\begin{equation}\label{eq:ocp_pareto}
	\begin{aligned}
		&  \max_{\bfU_{i}, i \in \mcalV}
		& & p \cdot \mathsf{W}_{developed} + (1-p) \cdot \mathsf{W}_{developing} \\
				&  {\rm subject \ to}
& & \bfx(t+1) = \bff(t, \bfx(t),\bfu(t)), \ \  \bfx(0) = \bfx_{0},  t \in \mcalT \\
&&& \bfu(t)   \in [0,1]^{24}, \ \  t \in \mcalT. 
	\end{aligned}
\end{equation}
where $p$ is selected in the interval $[0,1]$. For any fixed $p\in[0,1]$, we obtain a Pareto solution, and their collection form the Pareto frontier between the developed and developing clusters. 

\subsection{Receding Horizon Solution to RICE Social Welfare Maximization}
The work of \cite{MPC-DICE} established a novel receding horizon solution to DICE, which provides robustness and computational efficiency compared to solving DICE in a long time horizon directly. We  extend the idea of \cite{MPC-DICE} to RICE. 

For the receding horizon approach, we denote the prediction horizon by $T_{rh}$ and the simulation horizon by $T_{sim}$. We introduce $l(t, \bfx(t),\bfu(t)) :=\sum_{i = 1}^{N} c_{i}\cdot \frac{g_{i}(C_{i}(t),L_{i}(t))}{(1+\rho_{i})^{5t}}$. We assume a full measurement of the estimate of the state $\bfx(t)$ is available at each time step $t \in \mcalT_{sim}:= \{0,1,\dots,T_{sim}\}$. We present  the receding horizon approximation of ~\eqref{eq:swm} in Algorithm~\ref{alg:rch}:
	\begin{algorithm}
	\caption{MPC-RICE}\label{alg:rch}
	{\bf{Input:}} simulation horizon $T_{sim}$; prediction horizon $T_{rh}$.  \\
	\begin{algorithmic}[1]
		\State $t \gets 0$
		\While{\texttt{$t \leq T_{sim}$} }
		\State	{\bf observe} $\bfx(t)$ 
		
		\State	{\bf compute} the optimal solution ${\bfu^{\ast}(s),s \in \mathcal{S}:=\{t, t+1, \dots, t+T_{rh}\}},$ to the following optimization problem over the receding horizon $\mathcal{S}$
		\begin{equation}\label{eq:drh}
			\begin{aligned}
				&\max_{ \mathbf{u}(s),\forall s \in \mcalS} \quad 	& &  \sum_{s = t}^{t+T_{rh}} l(s, \bfx(s),\bfu(s)) \\
				&  {\rm subject \ to}
				& & \bfx(s+1) = \bff(s, \bfx(s),\bfu(s)),\ \  s \in \mathcal{S},\\
				&&& \bfu(s) \in [0,1]^{24}, \ \  s \in \mathcal{S}.
			\end{aligned}
		\end{equation}
		
		\State	{\bf apply} $\bfu^{\rm rhw}(t):=\bfu^{\ast}(t)$ to RICE dynamic game
		\EndWhile
	\end{algorithmic}
	\end{algorithm}

After Algorithm~\ref{alg:rch}, the control decision profile $\bfU^{\rm rhw} :=  [\bfu^{\rm rhw}(0)^{\top},\dots,\bfu^{\rm rhw}(T_{sim})^{\top}]^{\top}$ is the receding horizon solution to RICE.
\medskip

\section{Non-cooperative Decisions of the RICE Game}
In this section, we study the solutions to the RICE game under non-cooperative settings. We present the best-response dynamics for dynamic games, and receding horizon feedback decisions for dynamic games. It can be also implemented over the RICE game, since the RICE game is inherently a dynamic game.
\subsection{Best-response Dynamics}
In what follows, we establish the best-response recursions  for the dynamic   game introduced in Section~\ref{sec:DTDG} with $n$ players over a    finite horizon $T$. We assume the dynamic game is repeatedly and recursively  played for $N$ episodes, where each episode consists of $T$ time steps. We thereby define the aggregated control decisions of all players in episode  $k = 1,\cdots,N $   by $\bfU^{(k)}$, and the  decisions of player $i$ in episode  $k$ by $\bfU_{i}^{(k)}$. Similarly, the decisions of players excluding player $i$ in episode  $k$ is denoted by  $\bfU_{-i}^{(k)}$. The best-response recursion of the agents in the dynamic game over the $N$ episodes are described in  the following  Recursive Best-response Algorithm for Dynamic Games (RBA-DG) as in Algorithm~\ref{alg:dbr}. 

\begin{algorithm}[h]
	\caption{Recursive Best-response Algorithm for Dynamic Games (RBA-DG)}\label{alg:dbr}
	{\bf{Input:}} Episodes $N$; $c_{i}, i \in \mcalV$ .
	\begin{algorithmic}[1]
		\State  {\bf compute} an optimal cooperative solution $\bfU^{\rm c}$ by the following problem
		\begin{equation*}\label{eq:initialize}
			\begin{aligned}
				&  \max_{\bfU_{1},\cdots,\bfU_{n}}
				& & \sum_{i \in \mcalV} c_{i}\cdot J_{i}(\bfX,\bfU_{i},\bfU_{-i}) \\
				&  {\rm subject \ to}
				& & \bfx(t+1) = \bff(t, \bfx(t),\bfu(t)), \ \  \bfx(0) = \bfx_{0},  t \in \mcalT \\
				&&& \bfu(t)   \in [0,1]^{24}, \ \  t \in \mcalT. 
			\end{aligned}
		\end{equation*}
		\State {\bf let} $\bfU^{(0)}_{i} = \bfU^{\rm c}_{i}, \forall i  \in \mcalV$ 
		\State $k \gets 0$
		\While{\texttt{ k $\leq$ N} }
		\For{\texttt{each player $i \in \mcalV$}}
		\State {\bf observe} $\bfU^{(k)}_{-i}$ 
		\State {\bf compute} $\bfU^{(k+1)}_{i}$  by solving the problem
		\begin{subequations}\label{eq:dbr}
			\begin{align}
				\max_{\bfU_{i}}  & \quad \mathsf{J}_i(\bfX,\bfU_{i},\bfU_{-i}),\\
				s.t. &\quad 	\bfx(t+1) = \bff(t, \bfx(t),\bfu(t)),\\
				& \quad \bfU_{-i} = \bfU^{(k)}_{-i},\\
				& \quad \bfx(0) = \bfx_{0}.
			\end{align}
		\end{subequations}
		\EndFor
		\State $\bfU^{(k+1)} = [\bfU^{(k+1)}_{1};\cdots;\bfU^{(k+1)}_{N}]$
		\State $k \gets k+1$  
		\EndWhile
	\end{algorithmic}
\end{algorithm}

\subsection{Receding Horizon Feedback Decisions}
In what follows, we present a framework for dynamic games where in a single play over the time horizon,  players observe the underlying dynamic process and other players' actions, and then apply a receding-horizon feedback decision making   for a prediction horizon.

Consider  the dynamic   game introduced in Section~\ref{sec:DTDG} with $n$ players over a    finite horizon $T$. The game is played only once, and the players take the following feedback decision process in the receding horizon sense. The players apply a receding horizon approach and compute their feedback decisions $\mathbf{u}_i(t)$. At each time $t=0,\dots,T-1$, each player $i$ observes other players'  played action   $\mathbf{u}_{-i}(t)$ and the system state $\mathbf{x}(t)$. Then, every player $i$ assumes that $\mathbf{u}_{-i}(t)$ will continue to be played over $[t+1,t+T_{rh}]$, and therefore decides its best feedback decision plan $\mathbf{u}_{i|t+1\to t+T_{rh}}^{\rm RHP} (\mathbf{x}(t),\mathbf{u}_{-i}(t))$, where 
$\mathbf{u}_{i|t+1\to t+T_{rh}}^{\rm RHP}$ maximizes the cumulative payoff of player $i$ over the time horizon  $[t+1,t+T_{rh}]$ conditioned on that $\mathbf{u}_{-i}(t)$  be played over $[t+1,t+T_{rh}]$. Finally, each player $i$ plays the first planned decision $\mathbf{u}_{i|t+1}^{\rm RHP}$ in $\mathbf{u}_{i|t+1\to t+T_{rh}}^{\rm RHP}$ for the step $t+1$, and the process moves forward recursively.  Denoting $\mathbf{u}_i^{\rm RHF}(t)$ as the actions  generated by the receding horizon feedback decision process, clearly there is an underlying feedback law $\pi_i$ such that
$$
\mathbf{u}_i^{\rm RHF}(t)=\pi_i(t,\mathbf{x}(t),\mathbf{u}_{-i}^{\rm RHF}(t))
$$
which are actually played at each time $t$. 
The resulting  collective player decisions over the entire time horizon is written as $\mathbf{U}^{\rm RHF}$. The exact computational process of this receding horizon feedback decision framework is presented in the following  Receding Horizon Feedback Algorithm for Dynamic Games (RHFA-DG). 

\begin{algorithm}[h]
	\caption{Receding Horizon Feedback Algorithm for Dynamic Games (RHFA-DG)}\label{alg:drh}
	
	{\bf{Input:}} simulation horizon $T_{sim}$; prediction horizon $T_{rh}$; $c_{i}, i \in \mcalV$. 
	
	\begin{algorithmic}[1]
		\State {\bf compute} an optimal cooperative solution $\bfU^{\rm c}$ by the following problem
		\begin{equation*}\label{eq:initialize}
			\begin{aligned}
				&  \max_{\bfU_{1},\cdots,\bfU_{n}}
				& & \sum_{i \in \mcalV} c_{i}\cdot J_{i}(\bfX,\bfU_{i},\bfU_{-i}) \\
				&  {\rm subject \ to}
				& & \bfx(t+1) = \bff(t, \bfx(t),\bfu(t)), \ \  \bfx(0) = \bfx_{0},  t \in \mcalT \\
				&&& \bfu(t)   \in [0,1]^{24}, \ \  t \in \mcalT. 
			\end{aligned}
		\end{equation*}
		\State {\bf let} $\bfU^{\rm RHF}_{i}(0) = \bfU^{\rm c}_{i}(0), \forall i  \in \mcalV$ 
		
		\State $t \gets 0$
		
		\While{\texttt{$t \leq T_{sim}$} }
		\For{\texttt{each player $i \in \mcalV$}} 
		\State 	{\bf Take action}  $\bfU^{\rm RHF}_{i}(t)$
		\EndFor
		
		\For{\texttt{each player $i \in \mcalV$}} 
		\State {\bf observe} $\bfx(t)$ and $\bfU^{\rm RHF}(t)$
		
		\State {\bf compute} $\bfx(t+1)$ according to~\eqref{eq:DTDG_dynamics} 
		
		\State {\bf assume} all players $j \in \mcalV/\{i\}$ will continue to play $\bfU^{\rm RHF}_{-i}(t)$ over $[t+1,t+T_{rh}]$
		
		\State {\bf compute} its optimal solution $\bfu_{i|t+1\to t+T_{rh}}^{\rm RHP}$  to the following receding horizon optimization problem
		\begin{subequations}\label{eq:drh}
			\begin{align}
				\max_{ \bfu_{i|t+1\to t+T_{rh}}} \quad & \sum_{s = t+1}^{t+T_{rh}} g_{i}(s, \bfx(s),\bfu_{i}(s),\bfu_{-i}(s)) \\
				s.t. \quad & \bfx(s+1) = \bff(s, \bfx(s),\bfu(s)),\\
				& \bfu_{-i}(s) = \bfU^{\rm RHF}_{-i}(t),  \ \ s \in [t+1,t+T_{rh}].
			\end{align}
		\end{subequations}

		\State {\bf plan} $\bfU^{\rm RHF}_{i}(t+1) = \bfu_{i|t+1}^{\rm RHP}$
		\EndFor
		\EndWhile
	\end{algorithmic}
\end{algorithm}

\section{Description of Code}
In what follows, a description of how to use the repository {\RICEGAME} on GitHub is provided \cite{RICE-GAME}.  The repository \RICEGAME is a Matlab and CasADi-based implementation of the RICE game and its cooperative and non-cooperative solutions. A Matlab implementation of DICE and receding horizon solution to DICE can be found in \cite{DICEmc,MPC-DICE}.

The repository \RICEGAME consists of two folders:
\begin{itemize}
	\item \directory{/RICE-GAME/cooperative} provides cooperative solutions to the RICE game under cooperative settings;
	
	\item \directory{/RICE-GAME/non-cooperative} provides non-cooperative solutions to the RICE game under non-cooperative settings.
\end{itemize}

\subsection{Cooperative Solutions}
Under cooperative settings, the following cooperative solutions are considered:
\begin{itemize}
	\item Solution to RICE social welfare maximization under the directory
	
	 \directory{/RICE-GAME/cooperative/RICE\ SWM};
	
	\item Pareto frontier between developed and developing regions under the directory
	
	 \directory{/RICE-GAME/cooperative/Pareto\  Frontier};
	
	\item Receding horizon solution to RICE social welfare maximization under the directory
	
	 \directory{/RICE-GAME/cooperative/MPC-RICE}.
\end{itemize}
\medskip

{\noindent \textbf{Solution to RICE Social Welfare Maximization}.} It consists of seven files and a folder:
\begin{itemize}
	\item \path{RICE_SWM.m} is the top-level file and calls the files of \path{specify_parameters.m}, \path{solve_swm_problem.m}, \path{test_rice_dynamics.m}, \path{rice_dynamics.m}, and \path{try_a_guess.m}. It provides an implementation of RICE social welfare maximization.
	
	\item \path{specify_parameters.m} specifies and returns the parameters and exogenous terms for the RICE game, in the structure \path{Params}.
	
	\item \path{solve_swm_problem.m} is a function that solves RICE social welfare maximization problem~\eqref{eq:swm} given initial condition \path{x0} and time horizon \path{problem_ horizon}. 
	
	\item \path{test_rice_dynamics.m} is a function that takes all players' control decisions of \path{double} data type as one of input arguments and calculates the dynamical states of the RICE game (all dynamical states belong to \path{double} data type). In addition, it also calculates the value of the payoff functions (belonging to \path{double} data type) and the quantities for each region's emissions and consumption (belonging to \path{double} data type).
	
	\item \path{rice_dynamics.m} is a function that takes all players' control decisions of \path{SX} data type from CasADi as one of input arguments and calculates the dynamical states of the RICE game (all dynamical states belong to \path{SX} data type). In addition, it also calculates the value of the payoff functions (belonging to \path{SX} data type) and the quantities for each region's emissions and consumption (belonging to \path{SX} data type).
	
	\item \path{try_a_guess.m} returns an initial guess as a starting point for RICE social welfare maximization problem~\eqref{eq:swm}.
	
	\item \path{plot_result.m} generates plots of the trajectories of the emission-reduction rate, saving rate, atmospheric temperature deviation and social cost of $\rm CO_{2}$.
	
	\item \directory{./results} is a folder where the results computed from \path{RICE_SWM.m} are saved.
\end{itemize}
\medskip

{\noindent \textbf{Pareto Frontier between Developed and Developing Regions}.} It consists of five files and a folder:
\begin{itemize}
	\item \path{Pareto.m} is the top-level file and calls the files of \path{specify_parameters.m}, \path{rice_dynamics.m} and  \path{test_rice_dynamics.m}. It takes  $1001$ linearly spaced values between $0$ and $1$ as the values of $p$, and solves the respective optimization problem~\eqref{eq:ocp_pareto} under each $p$.
	
	\item \path{specify_parameters.m}, \path{rice_dynamics.m} and  \path{test_rice_dynamics.m} are the same as those in RICE social welfare maximization.
	
	\item \path{plot_pareto.m} generates plots of the social welfare Pareto frontier between developed regions and developing regions, and the atmospheric temperature deviation at the final time step versus the parameter $p$.
	
	\item \directory{./results} is a folder where the results computed from \path{Pareto.m} are saved. 
\end{itemize}

{\noindent \textbf{Receding Horizon Solution to RICE Social Welfare Maximization}.} It consists of seven files and a folder:
\begin{itemize}
	\item \path{MPC_RICE.m} is the top-level file and calls the files of \path{specify_parameters.m}, \path{solve_swm_problem.m}, \path{rice_dynamics.m}, \path{test_rice_dynamics.m} and \path{try_a_guess.m}. It provides an implementation of Algorithm~\ref{alg:rch}. The simulation horizon is specified in \path{T_simulation}. The prediction horizon is set in \path{T_prediction} and \path{problem_horizon}.
	
	\item \path{specify_parameters.m}, \path{solve_swm_problem.m}, \path{rice_dynamics.m}, \path{test_rice_dynamics.m} and \path{try_a_guess.m} are the same as those in RICE social welfare maximization.
	
	\item \path{comparison_dg_mpc.m} generates the plots of  the comparison of the atmospheric temperature deviation trajectories and optimal emission-reduction rates under the optimal control decisions solved from~\eqref{eq:swm} and Algorithm~\ref{alg:rch} with different receding horizons.
	
	\item \directory{./results} is a folder where the results computed from \path{MPC_RICE.m} are saved. 
\end{itemize} 

\subsection{Non-cooperative Solutions}
 Under non-cooperative settings, the following planning processes are considered: 
 \begin{itemize}
 	\item Best-response dynamics under the directory
 	
 	 \directory{/RICE-GAME/non-cooperative/BR};
 	
 	\item Receding horizon feedback decisions under the directory
 	
 	 \directory{/RICE-GAME/non-cooperative/RHFA}.
 \end{itemize}
 
{\noindent \textbf{Best-response Dynamics}.} There are six files and a folder:
\begin{itemize}
	\item	\path{BR_Algorithm.m} is the top-level file and calls the  files of \path{specify_parameters.m}, \path{test_rice_dynamics.m}, \path{ith_rice_dynamics.m} and \path{solve_ith_problem.m}. It provides an implementation of Algorithm~\ref{alg:dbr}.
	
	\item	\path{specify_parameters.m} and \path{test_rice_dynamics.m} are the same as those in RICE social welfare maximization.
	
	\item	\path{ith_rice_dynamics.m} is a function that takes player $i$'s control decisions of \path{SX} data type from CasADi and other players' control decisions of \path{double} data type as two of input arguments and calculates the dynamical states of the RICE game (all dynamical states belong to \path{SX} data type). In addition, it also calculates the value of the payoff functions (belonging to \path{SX} data type) and the quantities for each region's emissions and consumption (belonging to \path{SX} data type).

	\item	\path{solve_ith_problem.m} is a function that solves the optimization problem~\eqref{eq:dbr} in the iteration $k+1$ of the best-response dynamics given initial condition \path{x0}, time horizon \path{problem_horizon} and other players' control decisions \path{U_br(:,:,k)} in the iteration $k$.
	
	\item	\path{comparison_br.m} generates the plots of the convergence of $\|\bfU^{(k)}_{i} - \bfU^{(k-1)}_{i}\|$ versus iterations, and the comparison of the atmospheric temperature deviation trajectories and optimal emission-reduction rates under the optimal control decisions solved from~\eqref{eq:swm} and Algorithm~\ref{alg:dbr} in the final iteration.
	
	\item \directory{./results} is a folder where the results computed from \path{BR_Algorithm.m} are saved. 
\end{itemize}
\medskip

{\noindent \textbf{Receding Horizon Feedback Decisions}.} There are six files and a folder:
\begin{itemize}
	\item	\path{RHF_Algorithm.m} is the top-level file and calls the files of	\path{specify_parameters.m}, \path{test_rice_dynamics.m}, \path{solve_ith_problem.m} and \path{ith_rice_dynamics.m}. It provides an implementation of Algorithm~\ref{alg:drh}.
	
	\item	\path{specify_parameters.m} and \path{test_rice_dynamics.m} are the same as those in RICE social welfare maximization
	
	\item	\path{solve_ith_problem.m} and \path{ith_rice_dynamics.m} are the same as those in the best-response dynamics.
	
	\item	\path{comparison_RHFA.m} generates the plots of  the comparison of the atmospheric temperature deviation trajectories and optimal emission-reduction rates under the optimal control decisions solved from~\eqref{eq:swm}, Algorithm~\ref{alg:dbr} in the final iteration $K$, and Algorithm~\ref{alg:drh} with different prediction horizons.
	
	\item \directory{./results} is a folder where the results computed from \path{RHF_Algorithm.m} are saved. 
\end{itemize}

\section{Software Requirements}
This implementation of \RICEGAME is implemented in the platform of Matlab \cite{MatLab} along with the CasADi framework for automatic differentiation and numeric optimization \cite{CasADi}. Version 3.5.5 of CasADi is used. The compatible Matlab version for different operating systems can be found on the CasADi website \cite{CasADiweb}. The repository \RICEGAME is distributed under the  GNU General Public License v3.0.

\begin{table}[htbp] 
	
	\caption{Variables definitions for the RICE game. State variables and control variables are marked with asterisks $\ast$ and stars $\star$, respectively.}
	\centering
	\begin{tabular}{l|l|l}
		\hline
		Variable & Definition & Unit   \\ 
		\hline
		\multicolumn{3}{l}{Regions:} \\
		\hline
		$i$ & Index of a region  &  -  \\  
		\hline
		\multicolumn{3}{l}{Time steps and Calendar years:} \\
		\hline
		$t$ & Discrete time step  & -  \\ 
		$year(t)$ & Calendar year  &  -   \\ 
		\hline
		\multicolumn{3}{l}{Temperature dynamics:} \\
		\hline
		$^{\ast} T^{\rm AT}(t)$  & Atmospheric temperature deviation from the reference year $1750$  &  \textdegree{}C   \\ 
		$^{\ast} T^{\rm LO}(t)$  & Temperature deviation in lower ocean from the reference year $1750$  &  \textdegree{}C   \\ 
		$F(t)$  & Total radiative forcing  &  $\rm W/m^{2}$  \\ 
		$F^{\rm EX}(t)$  & Radiative forcing caused by greenhouse gases other than $\rm CO_{2}$  &  $\rm W/m^{2}$   \\ 
		\hline
		\multicolumn{3}{l}{Carbon dynamics:} \\
		\hline
		$^{\ast} M^{\rm AT}(t)$  & Carbon mass in reservoir for atmosphere  &  GtC  \\ 
		$^{\ast} M^{\rm UP}(t)$  & Carbon mass in reservoir for upper ocean  & GtC   \\ 
		$^{\ast} M^{\rm LO}(t)$  & Carbon mass in reservoir for lower ocean  & GtC  \\ 
		$\sigma_{i}(t)$ & Region $i$'s ratio of uncontrolled industrial emissions to gross economic output  &  GtC/trillions USD \\ 
		$^{\star} \mu_{i}(t)$ & Region $i$'s emission-reduction rate  &  -  \\ 
		$E_{i}^{\rm land}(t)$  & Region $i$'s natural $\rm CO_{2}$ emissions from land use   &  Gt$\rm CO_{2}$  \\ 
		$E_{i}(t)$  & Region $i$'s $\rm CO_{2}$ emissions including industrial emissions and natural emissions & Gt$\rm CO_{2}$  \\ 
		$E(t)$  & Global $\rm CO_{2}$ emissions across all regions   &  Gt$\rm CO_{2}$   \\ 
		\hline
		\multicolumn{3}{l}{Economic dynamics:} \\
		\hline
		$L_{i}(t)$ & Region $i$'s population  &  millions people  \\  
		$A_{i}(t)$ & Region $i$'s total productivity factor  &  -   \\ 
		$^{\ast} K_{i}(t)$ & Region $i$'s capital  &   trillions USD \\  
		$g_{i}(t)$ & Region $i$'s utility  &   trillions USD  \\ 
		$C_{i}(t)$ & Region $i$'s consumption  &   trillions USD \\ 
		$\mathsf{J}_{i}$ & Region $i$'s cumulative social welfare across the time horizon  &   trillions USD   \\ 
		$^{\star} s_{i}(t)$ & Saving rate  &  -  \\ 
		\hline
	\end{tabular}
	\label{table_variables}	
\end{table}

\begin{table}[htbp]
	\caption{Parameters for the RICE game.}
	\centering
	\begin{tabular}{l|l|l}
		\hline
		Parameter &  Description & Value  \\
		\hline
		\multicolumn{3}{l}{Regions:}   \\
		\hline
		$n$ & Number of regions & 12\\
		\hline
		\multicolumn{3}{l}{Time steps and Calendar years:}   \\
		\hline
		$\Delta$ & Sampling rate & 5 \\
		$T$ & Horizon length& 120\\
		$year(0)$ & Starting year& 2020\\
		\hline
		\multicolumn{3}{l}{Temperature dynamics:}   \\
		\hline
		$\phi_{11}$ & Diffusion coefficient between temperature layers  & 0.871810629   \\ 
		$\phi_{12}$ & Diffusion coefficient between temperature layers & 0.008844   \\
		$\phi_{21}$ & Diffusion coefficient between temperature layers & 0.025   \\
		$\phi_{22}$ & Diffusion coefficient between temperature layers & 0.975   \\
		$\eta$ & Forcing associated with equilibrium of carbon doubling (W/$\rm m^2$)& 3.6813  \\
		$\xi_{2}$   & Multiplier for $\eta$ & 0.1005 \\
		$M_{\rm AT,1750}$  & Atmospheric mass of carbon in the year $1750$ (GtC) & 588 \\
		\hline
		\multicolumn{3}{l}{Carbon dynamics:}   \\
		\hline
		$\zeta_{11}$ & Diffusion coefficient between carbon reservoirs  & 0.88  \\
		$\zeta_{12}$ &  Diffusion coefficient between carbon reservoirs & 0.196  \\
		$\zeta_{21}$ &  Diffusion coefficient between carbon reservoirs & 0.12  \\
		$\zeta_{22}$ & Diffusion coefficient between carbon reservoirs  & 0.797  \\
		$\zeta_{23}$ &  Diffusion coefficient between carbon reservoirs & 0.001465  \\
		$\zeta_{32}$ &  Diffusion coefficient between carbon reservoirs & 0.007  \\
		$\zeta_{33}$  & Diffusion coefficient between carbon reservoirs & 0.99853488  \\
		$\xi_{1}$  & Conversion factor from emissions to carbon mass (GtC / Gt$\rm CO_2$)& 5*0.27272727 \\
		\hline
		\multicolumn{3}{l}{Economic dynamics:}   \\
		\hline
		$\delta_{i}^{\rm K}$ & Region $i$'s depreciation rate on capital per year &  \multirow{11}{*}{See Table~\ref{table_ec_parameters}}  \\
		$a_{i}^{[1]}$ & Region $i$'s damage coefficient on temperature &  \\
		$a_{i}^{[2]}$ & Region $i$'s damage exponent &  \\
		$a_{i}^{[3]}$ & Region $i$'s damage coefficient on temperature squared &  \\
		$\theta_{i}^{[2]}$ & Region $i$'s exponent of emission-reduction cost function &  \\
		$\gamma_{i}$ &  Region $i$'s capital elasticity in gross economic output &  \\
		$pb_{i}$ & Region $i$' backstop technology price in the year $2020$  (USD/t$\rm CO_{2}$)&  \\
		$\delta_{i}^{pb}$ & Region $i$'s decline rate of backstop cost &  \\
		$\alpha_{i}$ & Region $i$'s elasticity of marginal utility of consumption& \\
		$\rho_{i}$ & Region $i$'s rate of social time preference per year& \\
		$c_{i}$ & Region $i$'s Negishi parameter in RICE social welfare function& \\
		\hline
	\end{tabular}	
	\label{table_gp_parameters}	
\end{table}

\begin{table}[htbp]
	\caption{Parameter Values for the economic dynamics.}
	\centering
	\begin{tabular}{l|llllllllllll}
		\hline
		& US & EU & JN & RS & EUR & CN & IN & ME & AF & LA & OHI & OA\\
		\hline
		$\delta_{i}^{\rm K}$ & 0.1 & 0.1 & 0.1 & 0.1 & 0.1 & 0.1 & 0.1 & 0.1 & 0.1 & 0.1 & 0.1 & 0.1 \\
		$a_{i}^{[1]}$ & 0    &     0    &     0    &     0    &     0  &  0.0008  &  0.0044  &  0.0028  &  0.0034   & 0.0006    &     0  &  0.0018 \\
		$a_{i}^{[2]}$ & 0.0014  &  0.0016 &   0.0016  &  0.0011 &  0.0013  &  0.0013  &  0.0017  &  0.0016  &  0.0020  &  0.0014   &  0.0016  &  0.0017 \\
		$a_{i}^{[3]}$ & 2 & 2 & 2 & 2 & 2 & 2 & 2 & 2& 2 & 2 & 2 & 2 \\
		$\theta_{i}^{[2]}$ & 2.6 &  2.6 &  2.6 &  2.6&  2.6 &  2.6 &  2.6 &  2.6 &  2.6 &  2.6 &  2.6 &  2.6\\ 
		$\gamma_{i}$ & 0.141 & 0.159 & 0.162 & 0.115 & 0.130 & 0.126 & 0.169  & 0.159 & 0.198 & 0.135 & 0.156 & 0.173 \\
		$pb_{i}$  & 1051    &    1635    &    1635        & 701    &    701    &     817     &   1284   &     1167    &    1284     &   1518    &    1284     &   1401 \\
		$\delta_{i}^{pb}$  & 0.025 & 0.025 & 0.025 & 0.025 & 0.025 & 0.025& 0.025 & 0.025 & 0.025 & 0.025 & 0.025 & 0.025 \\
		$\alpha_{i}$  & 1.45 & 1.45 & 1.45 & 1.45 & 1.45 & 1.45& 1.45 & 1.45 & 1.45 & 1.45 & 1.45 & 1.45\\
		$\rho_{i}$  & 0.015 & 0.015 & 0.015 & 0.015 & 0.015 & 0.015 & 0.015 & 0.015 & 0.015 & 0.015 & 0.015 & 0.015\\
		$c_{i}$ & 0.2010  &  0.1030  &  0.1300  &  0.0300  &  0.0080  &  0.0040  &  0.0020  &  0.0156   & 0.0013  &  0.0157  &  0.1187  &  0.0031\\
		\hline
	\end{tabular}	
	\label{table_ec_parameters}	
\end{table}

\begin{table}[htbp]
	\caption{Default initial condition for the geophysical dynamics.}
	\centering
	\begin{tabular}{l|ll}
		\hline
		 & Value  &   \\
		\hline
		$T^{\rm AT}(0)$ & 1.15 &  \\
		$T^{\rm LO}(0)$ & 0.05 &  \\
		$M^{\rm AT}(0)$ & 979 &  \\
		$M^{\rm UP}(0)$ & 485 &  \\
		$M^{\rm LO}(0)$ & 1741 &  \\
		
		\hline
	\end{tabular}	
	\label{table_gp_initial_condition}	
\end{table}

\begin{table}[htbp]
	\caption{Default initial condition for the economic dynamics.}
	\centering
	\begin{tabular}{l|llllllllllll}
		\hline
		& US & EU & JN & RS & EUR & CN & IN & ME & AF & LA & OHI & OA\\
		\hline
		$K_{i}(0)$ & 36.59&37.11&9.60&4.96&2.61&28.47&11.94&14.46&6.81&17.49&11.61&11.09 \\
		\hline
	\end{tabular}	
	\label{table_ec_initial_condition}	
\end{table}

\end{document}